# Automatic Translation of Unstructured Requirements into Linear Temporal Logic through Large Language Models

Alexandra Newcomb
*Department of Electrical Engineering and Computer Science*
*Embry-Riddle Aeronautical University*
Daytona Beach, FL, USA
davidofa@my.erau.edu

Omar Ochoa
*Department of Electrical Engineering and Computer Science*
*Embry-Riddle Aeronautical University*
Daytona Beach, FL, USA
ochoao@erau.edu

***Abstract*—Automatically translating unstructured natural language requirements into formal specifications remains a challenge in requirements engineering and formal methods, particularly for safety- and mission-critical systems whose verification depends on mathematically precise specifications. This paper evaluates whether contemporary off-the-shelf Large Language Models (LLMs) can help bridge this gap by generating Linear Temporal Logic (LTL) formulas directly from unstructured requirements. The study examines six modern LLMs using a few-shot prompting strategy on a heterogeneous benchmark of 15 structurally varied requirements. Five independent generations were collected for each requirement-model pair, yielding 450 candidate LTL formulas in total. Performance was assessed through manual semantic evaluation, pass@k for k ∈ {1, 3, 5}, and a self-consistency measure capturing syntactic reproducibility across stochastic trials. The results indicate that current general-purpose LLMs can achieve practically significant performance on the unstructured NL-to-LTL task without task-specific fine-tuning. The study also considers understandability for non-experts by pairing generated formulas with model-produced natural language explanations and discussing the complementary use of timeline-based LTL visualization. The findings suggest that modern LLMs are becoming viable front-end assistants for semi-automated formalization workflows.**

## I. INTRODUCTION

Automatically generating formal specifications from unstructured Natural Language (NL) requirements remains a significant challenge within the intersection of requirements engineering and formal methods [1, 2, 3]. Safety, security, and mission-critical software systems are often described initially in NL requirements, which are frequently both unstructured and ambiguous [2, 4, 5]. However, formal verification and synthesis techniques typically require mathematically precise specifications. Bridging this gap is necessary for enabling rigorous analysis early in the development lifecycle, for supporting formal verification later in the development lifecycle, and for reducing downstream rework caused by misinterpretation of requirements [6]. A core goal of research in this area is therefore to streamline the conversion of NL requirements into formal specifications and therefore broaden access to formal methods in industry [2, 3, 5].

At present, producing high-quality formal specifications from NL generally requires substantial expertise in formal methods [4, 5, 7]. This expertise is not only scarce but also costly to apply at scale, as manual formalization is both time-consuming and requires iterative clarification with domain stakeholders [2, 5]. As a result organizations either forego formal analysis entirely or restrict its use to a small subset of requirements where the cost can be justified [1, 7, 8]. Consequently, the formal methods community has sustained interest in automating or semi-automating this process, and current research explores a range of tools and methods aimed at the automatic formalization of requirements [1, 2, 6].

A significant portion of existing work focuses on translating structured NL into formal specifications [1, 2]. In these approaches, requirements are written using constrained templates, controlled vocabularies, or predefined patterns that reduce syntactic variability and limit semantic ambiguity [2, 6]. Such structure enables reliable mapping from structured NL phrases to formal operators and has led to practical tools that can generate temporal logic formulas, automata, or other machine-checkable artifacts [1]. However, structured requirements impose an additional burden on the software engineer in that the engineer must possess sufficient knowledge of the template language, the underlying formalization assumptions, and the boundary cases that cause template violations. In practice, this makes translating structured NL to formal specifications potentially costly to implement due to the additional training and experience required to correctly craft the structured NL requirements and interpret the resulting formal specifications.

In contrast, unstructured NL requirements are prevalent in industrial documentation but are inherently ambiguous and context dependent [2, 4, 6]. They often include underspecified conditions, implicit temporal assumptions, domain-specific terminology, and cross-references to external artifacts. These characteristics make unstructured requirements challenging to translate automatically into formal specifications using rule-based or template-driven pipelines, which typically depend on predictable syntactic forms. As a result, although structured-to-formal translation has seen notable progress, robust automation for unstructured-to-formal translation remains limited [1].

Large Language Models (LLMs) constitute a promising candidate technology for addressing this gap. Unlike earlier approaches that rely heavily on controlled input formats, LLMs are designed to process and generate free-form NL at scale.

Their training on broad corpora enables them to capture syntactic variation and to infer aspects of semantics that are not explicitly encoded in templates [1, 3]. Recent models of LLMs, including models tuned for reasoning, demonstrate strong performance on logical reasoning and software engineering tasks, suggesting potential utility in interpreting requirement statements and producing corresponding formal artifacts. These capabilities motivate the hypothesis that LLMs may reduce the need for structured requirement authoring while still enabling accurate generation of formal specifications.

The motivation of this paper is to evaluate how effectively modern off-the-shelf LLMs can automatically translate unstructured NL requirements into formal specifications expressed in Linear Temporal Logic (LTL). LTL is a widely used specification language for reactive systems and model checking, capable of expressing temporal properties such as safety (something bad never happens) and liveness (something good eventually happens) [9, 10, 11]. Because LTL properties are both expressive and support automated verification, successful translation from unstructured requirements to LTL would offer a direct pathway from informal documentation to machine-checkable specifications.

In addition to translation accuracy, a central challenge addressed in this study is evaluation by non-experts: even if an LLM produces syntactically valid LTL, determining whether the formula correctly matches the intended meaning of the original requirement is nontrivial without formal methods expertise. This study therefore also investigates techniques to make the generated outputs more understandable and verifiable by practitioners who are not specialists in temporal logic. Two complementary mechanisms are considered. First, the study prompts models to produce detailed yet brief explanations for each generated LTL formula, explicitly mapping parts of the NL requirement to components of the temporal logic expression. Second, the study briefly discusses ltl2timeline, which converts LTL formulas into timeline-style visualizations intended to expose the temporal structure of a property and support intuitive inspection [9]. Together, these techniques aim to support a workflow in which stakeholders can assess semantic alignment between requirements and generated LTL without needing to be experts in the specification language themselves.

The following research questions are posed in this study:

1. How well do the selected off-the-shelf LLMs perform on the NL-to-LTL translation task?
2. How consistently do the selected off-the-shelf LLMs output the same LTL formula, defined as syntactic and semantic equivalence?

By focusing on unstructured NL as the input modality and by emphasizing understandability and evaluation for non-experts, this work seeks to decrease the cost of expert-driven formalization and the difficulty of validating formal artifacts against stakeholder intent (in that the formal artifacts are traceable to NL requirements). The remainder of the paper investigates the extent to which contemporary LLMs can mitigate these barriers and outlines evaluation strategies designed to make formal specifications accessible to a wider engineering audience.

## II. Related Work

The automatic translation of NL requirements into formal specifications has seen increased interest as a means of reducing the cost and specialized expertise required to apply formal verification to real systems. Much of the established progress in this area has been achieved by constraining the input language through templates, controlled vocabularies, or structured requirement formats, which reduce syntactic variability and limit ambiguity, thereby enabling more reliable mappings from text to formal representations. In contrast, unrestricted (unstructured) NL remains difficult to formalize automatically due to ambiguity, domain-specific terminology, and substantial variation in phrasing [1, 2]. Recent work has increasingly investigated NLP and LLM-based approaches to better accommodate unstructured inputs while still producing machine-checkable temporal logic specifications suitable for downstream analysis [3, 12]. Currently, only a small number of tools accept unrestricted NL directly, a gap which this work addresses [1].

The Automatic Requirements Specification Extraction from Natural Language (ARSENAL) framework provides an early end-to-end framework for extracting formal models and temporal logic specifications from NL requirements [13]. The framework translates requirements into models expressed in the Symbolic Analysis Laboratory modeling language and specifications expressed in LTL. Separately, FRET focuses on translating structured requirements [1, 14]. FRET accepts requirements written in a structured language, FRETish, and translates them into past and future-time metric temporal logic formulas. Past-time metric temporal logic requirements are translated into Lustre, a synchronous dataflow language, and the resulting Lustre models are analyzed by backend engines for realizability. Future-time metric temporal logic requirements are formulated within a NuSMV model and analyzed using the NuSMV model checker [1, 14].

On the other hand, NL2CTL proposes an LLM-based framework that translates NL requirements into Computation Tree Logic (CTL), aiming to reduce the manual effort and expertise needed to produce specifications for model checking [3]. Because paired NL–CTL data are scarce, NL2CTL fine-tunes a T5-Large model on synthetically generated training data. The reported performance indicates that the task remains challenging: the fine-tuned approach achieves 46.4% accuracy, and an LLM baseline relying on few-shot learning and prompt engineering achieves only 2% accuracy in the evaluated setting. These results suggest that direct prompting alone may be insufficient for reliable temporal-logic synthesis from NL, and that even supervised approaches face difficulties related to semantic ambiguity and the precise structural constraints of branching-time logics [3].

NL2Spec addresses unstructured NL more directly through an interactive, human-in-the-loop translation framework that uses LLMs to produce temporal logic specifications and to expose intermediate rationale that supports traceability and correction [12]. Beyond generating a final LTL formula, NL2Spec prompts the model to construct a dictionary of sub-translations that map fragments of the requirement to fragments of the formal expression. This representation provides a form of explainability by making explicit which portions of the input correspond to particular operators and subformulas. The sub-translations support localized repair when the final formula is incorrect by allowing users to adjust individual sub-translations instead of rewriting an entire specification. The system takes as input an unstructured requirement and optionally user-provided or previously generated sub-translations. The backend then builds an interactive few-shot prompt, and the model produces a step-by-step NL explanation, the explanation dictionary, and the final LTL. This approach reflects a broader trend toward combining LLM generation with artifacts that enable auditing and iterative refinement [12].

Lastly, Leong and Barbosa investigate the translation of NL requirements into Java Modeling Language (JML) specifications, comparing a symbolic NLP pipeline to a GPT-based pipeline for this task [15]. The findings indicate that GPT often captures intended meaning but may generate JML that misuses Java or JML constraints unless guided by additional information.

Notably, although there is an increased interest in applying LLMs to the translation of NL requirements to formal specifications, little recent work has investigated the efficacy of the state-of-the-art LLMs for this task. This study evaluates six diverse LLMs with few-shot prompting on 15 different unstructured NL requirements, taking advantage of modern LLMs' strength at deconstructing underlying semantic structure while containing a significant general-purpose knowledge space. These combined strengths allow LLMs to generalize to tasks they were not explicitly trained on, such as the NL to formal specification task.

## III. Background

This section provides background information on LTL and the two metrics used during evaluation: pass@k and the self-consistency score.

### A. Linear Temporal Logic

LTL is a formal specification language for reasoning over properties of infinite sequences of system states (traces) by describing how propositions evolve over time along a system trace. LTL is used for translating specified safety properties and requirements into discrete mathematical formulas [11, 16]. The temporal operators used in this work are Eventually (**F**), Globally (**G**), and Until (**U**). LTL is an extension to propositional logic. Common propositional operators include: Or ( | ), And (&), If-then (->), and Not (!). Take the propositions *a* and *b*: **F***a* is true if there is a reachable future state in which a is true, and **G***a* is true if a is true in all future states. On the other hand, *a* **U** *b* is true if *a* is true at least until *b* becomes true.

The LTL formulas discussed in this work use the uppercase letter representations as opposed to the symbolic representations commonly used in the literature to support integration with the ltl2timeline tool, which expresses LTLs using ASCII text [9]. ASCII representations are also used to simplify instructions for LLM generations.

### B. Pass@k

The pass@k metric is a standard evaluation measure for generative systems that can produce multiple candidate outputs per input. As a result, pass@k is a popular metric for evaluating LLM generations. For a given requirement in this study, pass@k is the probability that at least one of the top *k* generated candidates is a correct LTL both syntactically and semantically [17]. In practice, pass@k is estimated by sampling *n* candidates from the model and checking how many are correct; if *c* of the *n* samples (for one problem) pass, then an unbiased estimator commonly used in the literature is shown in (1), with the convention that pass@k = 1 when $n - c < k$. This metric captures how quickly correctness is achieved as more attempts are allowed, reflecting both single-shot accuracy (small *k*) and the benefit of generating and validating multiple candidates (larger *k*) [17]. To capture the overall pass@k for a model, the pass@k for all problems are averaged.

$$pass@k = 1 - \frac{\binom{n-c}{k}}{\binom{n}{k}} \tag{1}$$

### C. Self-Consistency in LLMs

The self-consistency measure $C_{m,r}$ quantifies how reproducibly a model *m* generates the same LTL formula across repeated stochastic runs for the same requirement *r*. Put simply, this metric quantifies how consistently each LLM generates identical LTLs for the same requirement. The equation for this metric is shown in (2), where *K* represents the number of unique classes, $n_k$ is the number of generations belonging to that equivalence class, and *n* is the total number of trials for a given model and requirement.

$$C_{m,r} = \frac{\sum_{k=1}^{K}\binom{n_k}{2}}{\binom{n}{2}} = \frac{1}{n(n-1)}\sum_{k=1}^{K} n_k(n_k - 1) \tag{2}$$

For each requirement, the $n = 5$ trials are evaluated for equivalence using an equivalence relation (section IV discusses the equivalence relation used in more detail). If the 5 trials yield K unique formulas with multiplicities $n_1, \ldots, n_K$, self-consistency is defined as the fraction of agreeing trial pairs among all $\binom{5}{2} = 10$ pairs. Methodologically, $C_{m,r}$ is identical to the item-level observed agreement term $P_i$ in Fleiss' multi-rater agreement framework [18].

## IV. Approach

Fig. 1 summarizes the experimental setup. Fifteen natural-language requirements were selected from the PURE dataset as candidates for translation into LTL [19]. Selection prioritized requirements that inherently involve temporal reasoning over streams of states, as LTL semantics are defined over infinite traces [16]. To ensure that evaluation reflects the breadth of temporal constructs commonly used in practice, the chosen requirements were also curated to provide coverage of multiple recurring specification patterns, including safety-style invariants (e.g., "always"), liveness-style eventualities (e.g., "eventually"), and response/precedence relationships (e.g. "this until that") that naturally map to combinations of Globally (G), Eventually (F), and Until (U). To reduce the risk that findings are attributed to a single writing style or domain, the requirements were drawn from seven distinct Software Requirement Specification documents from the PURE dataset spanning both industrial and university settings. This sampling strategy yielded a small but deliberately heterogeneous benchmark intended to stress syntactic and semantic variation typical of unstructured requirements.

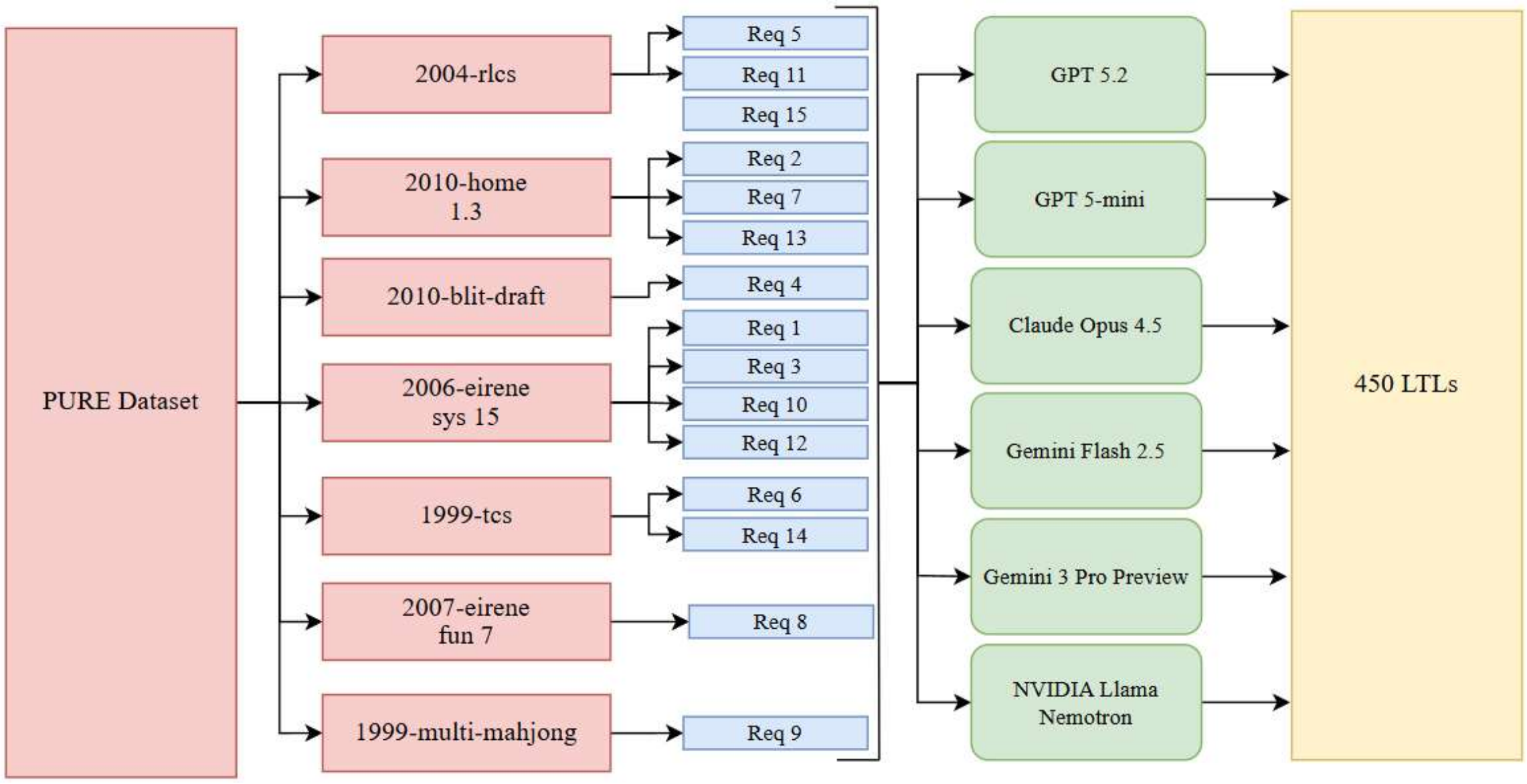

Fig. 1. Summary of experimental setup.

Each of the 15 requirements was manually formalized into LTL prior to data collection to provide a reference artifact during evaluation. Because unstructured NL often admits multiple correct formalizations (e.g., differences in scoping or alternative but logically equivalent encodings), these manual formulas were treated as guides rather than the unique ground truth. In particular, the manual formalizations were used to accelerate review by highlighting the expected temporal structure and key propositions, while the correctness decision for an LLM-generated formula was based on semantic and logical alignment with the requirement rather than exact string matching the reference.

Alongside each manual translation, a brief description of the components expected to appear in correct LLM outputs was noted. These component descriptions served as an evaluation checklist capturing (i) the intended atomic propositions (signals/events/states) referenced in the requirement, (ii) the required temporal operators and their scope (e.g., whether a constraint must hold globally or only after a trigger), and (iii) any necessary implication or ordering relationship between conditions and responses. The 15 selected NL requirements and their corresponding human-written LTL translations are provided in the supplemental data (see section VIII).

Table I summarizes the 15 selected requirements. Specifically, a brief description characterizing the translation task and the estimated difficulty of the task for each requirement is provided. Note that of the 15 requirements, three are classified as "easy," four are "medium," five are "hard," and three are "very hard." A range of difficulties was chosen to evaluate how well the different LLMs performed on increasingly challenging LTL translation tasks. While it is hypothesized that LLMs perform well on easy or medium tasks, it is possible that they will not perform well on the hard or very hard requirements. It is especially of interest to determine at what difficulty the LLMs are no longer able to reliably produce correct translations.

To evaluate how model choice impacts translation quality and cost-effectiveness, six LLMs were selected for comparison. The set was constructed to vary in model family, scale, and expected inference cost, which enables analysis of whether accuracy improvements (if any) justify the operational expense of using larger models. Special emphasis was placed on smaller and more economical models—such as NVIDIA Llama, Gemini Flash, and GPT 5-mini—because these models are potentially more plausible candidates for integration into engineering workflows where repeated querying is required. Additionally, the ability to run certain models locally, specifically Llama, is practically relevant for organizations with strict data governance requirements. Therefore, it is of interest to evaluate a diverse set of LLMs. The LLMs chosen for this study are:

1. OpenAI: GPT 5.2 [20]
2. OpenAI: GPT 5-mini [21]
3. Anthropic: Claude Opus 4.5 [22]
4. Google: Gemini 2.5 Flash [23]
5. Google: Gemini 3 Pro Preview [24]
6. NVIDIA: Llama 3.3 Nemotron Super 49B v1.5 [25]

Few-shot prompting was used during data collection, which exposes an LLM with a few input-output examples to guide task completion [26]. As a result, few-shot prompting provides a model-agnostic method for conditioning outputs toward a target syntax (ASCII LTL) and task framing (translation and brief justification), thereby preserving comparability across different LLMs. The few-shot prompt included two examples, each consisting of an NL requirement paired with an LTL formula written using the ASCII operator conventions (e.g., G, F, U, and propositional connectives). In addition to requesting an LTL formula, the prompt explicitly instructed the model to produce a brief explanation mapping subphrases in the requirement to corresponding subformulas and temporal operators. This explanation requirement was included to support understandability and to provide additional information during manual evaluation when a formula's intent was unclear from syntax alone.

The primary purpose of the natural language explanation of each translated LTL by the LLM is to provide a clear and concise explanation of the LTL mapping to developers that are non-experts in formal methods. This supports a key goal of this work, which is to investigate methods of utilizing LLMs to increase the adoption of formal methods in industry. As a result, these generated explanations can be used by the developer to both understand the generated LTL syntax and semantics and to validate that the LLM's specific interpretation of the NL requirement, and therefore the assumptions used for the LTL translation, are correct. The data collection prompt is provided in the supplemental data (see section VIII).

TABLE I. SUMMARY OF REQUIREMENT DIFFICULTY AND CHARACTERISTICS.

| Req ID | Difficulty | LTL Characteristics |
|---|---|---|
| 1 | Easy | Simple implication |
| 2 | Easy | Simple global condition |
| 3 | Easy | Simple until condition |
| 4 | Medium | Implication and X or F |
| 5 | Medium | Global condition with multiple propositions |
| 6 | Medium | Multiple propositional operators and globally condition |
| 7 | Medium | Requires temporal reasoning about ambiguous requirement |
| 8 | Hard | Requires reasoning about temporal conditions and user roles |
| 9 | Hard | Requires temporal reasoning for sequential states |
| 10 | Hard | Until condition with multiple propositional operators |
| 11 | Hard | Difficult temporal reasoning and ambiguous requirement |
| 12 | Hard | Many propositional statements |
| 13 | Very Hard | Multiple challenging and connected logical components |
| 14 | Very Hard | Challenging temporal reasoning |
| 15 | Very Hard | Challenging temporal reasoning |

For each combination of requirement and LLM, five independent trials were executed, yielding 15 requirements × 6 LLMs × 5 trials = 450 generated LTL formulas. Trials were performed independently to measure both achievable correctness under multiple attempts and the consistency of model outputs under stochastic generation. Independence was enforced by running each trial as a separate session with no cross-trial conversational context, preventing earlier outputs from influencing later ones.

The resulting corpus of 450 formulas was then manually evaluated for correctness. Logically equivalent formulations to the human-made LTL translations were accepted, consistent with the premise that multiple LTL encodings may capture the same requirement intent. Each generated LTL was inspected to determine whether it correctly represented the meaning of the originating unstructured requirement, with attention to:

1. Correct identification of triggers and responses,
2. Correct temporal scoping (e.g. whether a constraint is required globally or only under specific conditions),
3. Well-formed use of atomic propositions and operators, and
4. Correct logical outcomes (i.e. if the LTL correctly evaluates to true or false based on the expected outcome implied by the NL requirement)

The evaluation combines quantitative and qualitative analyses. Quantitatively, performance is summarized using pass@k computed from the five trials per requirement, characterizing the probability that at least one of the top $k$ samples constitutes a correct translation. Values chosen for $k$ are 1, 3, and 5.

In parallel, a self-consistency score was computed to quantify how reproducibly each model generated the same LTL formula across repeated stochastic trials for a given requirement. For each requirement, the five generated formulas were compared pairwise, and self-consistency was defined as the fraction of trial pairs that agreed under the equivalence relation used in this study. Two LTL formulas were considered equivalent if they were syntactically identical, allowing only for variation in proposition symbols and for differences in parenthesis that did not affect the formula structure. This allowance was necessary because generations were produced in independent sessions, in which different proposition letters were often assigned to the same requirement. Aside from these two exceptions, all aspects of the LTL structure had to match exactly for two formulas to be deemed equivalent.

Qualitatively, structured notes were recorded during manual review to document the reason for each recorded incorrect translation and to identify patterns in incorrect generations, such as missing temporal operators, incorrect logical assumptions, incorrect usage of operators, or inadequate formulations of propositions. These notes were then analyzed across models and requirements to determine tendencies in LLM outputs.

## V. RESULTS

This section presents the results of the unstructured NL to LTL translation task across the 15 requirements and six LLMs. Quantitative and qualitative results are presented in respective sections.

### A. Quantitative Results

Table II presents the average pass@k for each requirement across all six LLMs, for $k = 1, 3,$ and 5. Pass@1 represents the average success rate for a given requirement. Pass@3 represents the rate at which at least one of three sampled outputs (out of five total outputs) is correct. Pass@5 represents the rate at which at least one of the five generated LTL translations correctly

TABLE II. AVERAGE PASS@K VALUES FOR $K = 1, 3,$ AND 5 PER REQUIREMENT.

| Req ID | Pass@1 | Pass@3 | Pass@5 |
|---|---|---|---|
| 1 | 1.00 | 1.00 | 1.00 |
| 2 | 0.97 | 1.00 | 1.00 |
| 3 | 0.97 | 1.00 | 1.00 |
| 4 | 0.63 | 0.93 | 1.00 |
| 5 | 0.90 | 0.98 | 1.00 |
| 6 | 0.17 | 0.27 | 0.33 |
| 7 | 0.87 | 1.00 | 1.00 |
| 8 | 0.27 | 0.42 | 0.50 |
| 9 | 0.63 | 0.87 | 1.00 |
| 10 | 1.00 | 1.00 | 1.00 |
| 11 | 0.97 | 1.00 | 1.00 |
| 12 | 0.87 | 0.93 | 1.00 |
| 13 | 0.83 | 1.00 | 1.00 |
| 14 | 0.67 | 0.97 | 1.00 |
| 15 | 0.47 | 0.73 | 0.83 |

matches the NL requirement. All pass@k values in Table III are averaged across the six LLMs.

Table III presents the average pass@k values for each LLM, averaged across each requirement. Values for *k* presented are similarly 1, 3, and 5.

To measure the consistency of model generations, the consistency metric introduced in section III.C was used. The results, presented in Table IV, indicate how frequently each LLM produced matching pairs of LTLs in independent sessions. These values are averaged across each requirement per LLM. A self-consistency score of 1 indicates that all generated LTLs were equivalent, whereas a self-consistency score of 0 indicates that each generated LTL for that LLM was different in structure.

### B. Qualitative Results

This section characterizes the results of unsuccessful LTL generations. Specifically, explanations for why some translations were marked as unsuccessful are listed below. Correct evaluations did not contain any of the listed errors. The number of errors belonging to each listed item is provided in parentheses. Each LTL designated as incorrect has at least one of the errors listed below. Note that some LTLs had multiple errors.

1. A proposition was written in the negative (count: 8).
2. Missing expected temporal operator (count: 18).
3. LTL results in a logical implication not originally implied by the requirement (count: 7).
4. Incorrect logic in that the resulting LTL may incorrectly evaluate to true under violating conditions (count: 23).
5. Missing logic from the NL requirement, such as an omission of one or more phrases from the requirement resulting in missing logic in the LTL (count: 32).
6. Defines propositions that are not used in the LTL (count: 12).
7. Incorrect formulation of a proposition (count: 29).
8. Incorrect usage of a propositional or temporal operator (count: 6).
9. Incorrect syntax used (count: 2).

## VI. Discussion

This section discusses the experimental results and provides recommendations for future use. Additionally, usage of LLMs alongside ltl2timeline for the translation task is described and the implications of this work are provided for the research community [9]. Table V summarizes the key findings and takeaways at the end of this section.

### A. Discussion of Quantitative Results

This section answers the two research questions concerning translation performance and LTL consistency. The quantitative results indicate that contemporary LLMs can perform the unstructured NL-to-LTL translation task with a level of accuracy that is already practically significant, particularly given that the models were tested with few-shot prompting only and without any task-specific fine-tuning. At the requirement level, Table II shows that a substantial portion of the benchmark was handled well across models. Reqs 1, 10, and, at the pass@5 level, most of the benchmark approached or reached perfect performance. The results suggests that once a requirement's semantic structure can be clearly decomposed into trigger, response, and temporal scope, modern LLMs often succeed in producing a correct formalization within a small number of attempts.

TABLE III. Average pass@k values for *k* = 1, 3, and 5 per LLM.

| LLM | Pass@1 | Pass@3 | Pass@5 |
|---|---|---|---|
| GPT 5.2 | 0.83 | 0.90 | 0.93 |
| GPT 5-mini | 0.63 | 0.81 | 0.87 |
| Claude Opus 4.5 | 0.75 | 0.84 | 0.87 |
| Gemini 2.5 Flash | 0.73 | 0.90 | 0.93 |
| Gemini 3 Pro Preview | 0.81 | 0.93 | 0.93 |
| Llama 3.3 Nemotron Super 49B v1.5 | 0.69 | 0.87 | 0.93 |

TABLE IV. Self-consistency score averaged across LLMs.

| LLM | Self-consistency Score |
|---|---|
| GPT 5.2 | 0.37 |
| GPT 5-mini | 0.43 |
| Claude Opus 4.5 | 0.62 |
| Gemini 2.5 Flash | 0.41 |
| Gemini 3 Pro Preview | 0.67 |
| Llama 3.3 Nemotron Super 49B v1.5 | 0.50 |

At the same time, the requirement-level results reveal that translation difficulty was not determined solely by the apparent complexity of the requirement text. The weakest results were obtained for Reqs 6 and 8. Req 6 achieved only 0.17 for pass@1, 0.27 for pass@3, and 0.33 for pass@5, while Req 8 achieved 0.27, 0.42, and 0.50, respectively. These are notably lower than the rest of the benchmark and indicate persistent failure modes rather than isolated mistakes.

Req 6 was especially informative because the low performance appears to stem from a recurring misunderstanding of implication. Across models, the generated LTLs often used the if-then operator in the wrong direction for Req 6. Although such formulas can appear plausible on first inspection, they are logically inadequate because they evaluate to true in cases where the underlying NL requirement is violated. A representative example generated by GPT-5mini for Req 6, was G((a & !e & !h) -> Xp), where 'a' denotes that an automatic data terminal control mode selection is active, 'e' denotes that EMCON is active, 'h' denotes that HERO is active, and 'p' denotes that manual override is permitted. The requirement states that manual override is allowed except during EMCON and HERO conditions, where the exact meaning of the acronyms is not relevant for LTL formulation. However, in the generated formula, if either 'e' or 'h' is true, then the antecedent (a & !e & !h) becomes false. Because an implication is false only when its left-hand side is true and its right-hand side is false, the full implication evaluates to true whenever the antecedent is false. As a result, the formula does not properly constrain the prohibited cases and can evaluate to true even when the intended requirement is not enforced. This error is subtle because the formula superficially resembles a correct condition-response pattern, yet its truth conditions do not align with the semantics of the requirement.

Req 8 exhibited a different but equally consistent failure mode. Req 8 specified "It shall only be possible for the user who initiated the call to talk, other users can only listen." Here, the generated LTLs often omitted logic for the listening user and instead focused only on the speaking user, or vice versa. The NL requirement expresses a role-sensitive exclusivity condition: the initiating user may talk, while all other users may only listen. Many generations captured only one side of that distinction. This indicates difficulty in preserving all semantically necessary actors when a requirement embeds contrasting permissions across different roles. The problem was therefore not primarily one of syntax, but of faithfully preserving the full relational structure of the requirement.

These two requirements illustrate a broader pattern visible across the benchmark: the more directly a requirement's subphrases can be mapped into LTL operators and atomic propositions, the higher the pass@k values were. Requirements with a clear condition-response or until-termination structure generally resulted in a higher success rate. By contrast, requirements that demanded additional reasoning about permission, exclusion, or role differentiation were substantially harder, even when they appeared short or simple in prose. This helps explain why Req 12, despite containing many propositions, was comparatively manageable: its internal structure aligns relatively well with LTL decomposition. In contrast, Req 14 appears simple at the surface level, yet requires a more careful reasoning process to infer the appropriate formal relation between message receipt, checking, error detection, and eventual correction.

This same observation helps explain the strong performance on Req 10. Although it was rated Hard, all models collectively achieved perfect scores on it in Table II. The likely explanation is that the requirement is structurally explicit. It presents a failed setup condition followed by a re-attempt behavior that persists until one of several stopping conditions occurs. Such phrasing aligns naturally with implication and until-based LTL patterns. Thus, benchmark difficulty as assigned qualitatively did not always correspond to effective difficulty for the models; structural regularity mattered more than surface complexity alone.

At the model level, the results in Table III show meaningful but not extreme separation among the six LLMs. GPT 5.2 achieved the highest pass@1 score at 0.83, indicating the strongest single-attempt reliability. Gemini 3 Pro Preview achieved the highest pass@3 score at 0.93, suggesting that it benefited most from multiple stochastic attempts. By pass@5, performance became relatively compressed: GPT 5.2, Gemini 2.5 Flash, Gemini 3 Pro Preview, and NVIDIA Llama all reached 0.93, while GPT 5-mini and Claude Opus 4.5 reached 0.87. This convergence is notable. Even NVIDIA Llama, which had the lowest pass@1 performance among the better-performing models except GPT 5-mini, was comparatively competitive once multiple samples were considered.

However, from an industrial perspective, pass@1 is likely the most relevant metric. A workflow that depends on generating several candidates and then selecting the best one assumes that a developer is available who can evaluate competing LTL formulations. In practice, that evaluator would likely need at least some familiarity with temporal logic to distinguish a semantically correct candidate from one that is merely plausible. Accordingly, higher pass@3 and pass@5 values are encouraging, but they do not fully offset weaker single-shot performance in realistic deployment contexts. On the other hand, a developer can generate multiple candidate LTLs and determine the best representation using the provided explanation and/or an LTL visualization tool, which is further explained in section VI.D.

The consistency results in Table IV provide an additional perspective on model performance. Gemini 3 Pro Preview and Claude Opus 4.5 were the most consistent generators, with self-consistency scores of 0.67 and 0.62, respectively. GPT 5.2 was the least consistent at 0.37. This is an interesting contrast, as GPT 5.2 also had the highest pass@1 score. The result suggests that lower consistency does not necessarily indicate worse performance overall. In several cases, GPT 5.2 produced distinct LTL formulas across multiple trials that were semantically equivalent but structurally different. Thus, inconsistency partly reflects expressive variation rather than simple instability. Even so, from a workflow standpoint, higher consistency may still be desirable because reproducible outputs are easier to audit, compare, and integrate into downstream artifacts.

Overall, the quantitative results support the conclusion that current LLMs are already largely successful at this translation task under lightweight prompting conditions. This finding is important because no fine-tuning, domain adaptation, or retrieval augmentation was used. Even ambiguous requirements such as Req 11, were often translated adequately and exhibited high pass@k performance. The results therefore suggest that the task is increasingly within reach for general-purpose LLMs, particularly if future usage incorporates stronger prompt design, better proposition control, and targeted human review of known failure modes.

### *B. Areas for Improvement*

A major area for improvement concerns the quality of the selected atomic propositions. In many successful outputs, the propositions were sufficiently clear to support a correct LTL translation. However, proposition formulation was also one of the most common sources of weakness, and the qualitative results indicate that proposition-related issues frequently accompanied otherwise promising formalizations.

Most notably, the models often tended to combine multiple behaviors or conditions into a single proposition rather than keeping propositions atomic. This reduces clarity and weakens downstream reusability. An atomic proposition should typically represent a single state, event, or condition that can be independently referenced elsewhere in the formal model. When a model collapses several behaviors into one proposition, the resulting LTL may still be interpretable for a single requirement, but the proposition set becomes less useful as part of a larger specification ecosystem.

The models also sometimes embedded logic directly inside propositions. For example, a proposition such as "the system state changes to…" or similarly transition-laden phrasing is unideal because it obscures whether the proposition denotes a state, an event, or an already-composed logical relation. This undermines the clean separation between atomic vocabulary and formula-level composition that formal specifications benefit from.

Another recurring issue was the treatment of negation. Rather than consistently defining propositions in positive form and using the negation operator ('!') in the LTL where needed, several generations introduced propositions that were themselves written negatively. Closely related to this, models rarely reused a positive proposition and simply negated it when the opposite condition was needed. Instead, models often created a second proposition to represent the opposite event or state. For example, instead of using a proposition for "connection successful" and negating it when an unsuccessful connection must be represented, some outputs introduced a separate proposition for "connection unsuccessful." This practice makes the proposition vocabulary larger, less coherent, and less reusable. It also increases the chance of semantic inconsistency if both positive and negative versions are treated as independent variables.

A further improvement concerns strict adherence to the ASCII syntax constraints provided in the prompt. Some models occasionally used symbolic logical notation instead of the requested ASCII operators, such as → instead of ->, ∧ instead of &, or ∨ instead of |. GPT 5.2 and Claude Opus 4.5 were the only models that did not exhibit this issue. Claude also consistently formatted propositions in a table rather than a list, despite the examples showing a list format. This formatting deviation was not significant. These issues were generally not counted against a generation unless a model used an entirely different logical operator that was not permitted. This occurred once with the use of if and only if, which was outside the allowed operator set and therefore represented a substantive deviation rather than a minor formatting inconsistency.

### C. *Recommendations for Future Use*

Several practical recommendations follow from these findings. First, prompts should be explicit about proposition design. In particular, instructions should clearly require propositions to be atomic. Although some outputs already exhibited this property, many generations simplified the requirement by embedding multiple system behaviors into a single proposition. That may appear convenient, but it reduces traceability and weakens reuse in larger formal models.

Proposition development would likely benefit from additional structure in the prompt or workflow. The results suggest that models should be guided to produce propositions that are positive, atomic, and reusable across a broader system model. Many generated propositions were overly specific to the local wording of an individual requirement and therefore less suitable for integration into a larger specification set. This is an important limitation because industrial formalization rarely ends with a single isolated formula; the long-term value of the proposition set depends heavily on whether it can be reused coherently across related requirements. The models were not explicitly prompted with this systems-level objective, and the proposition quality reflects that omission.

Accuracy may improve if the propositions are supplied to the model in advance rather than generated jointly with the LTL. Doing so would reduce one major source of variability and force the translation task to focus on temporal and logical composition. The tradeoff, however, is that this approach imposes an additional burden on the developer, who must define the proposition vocabulary beforehand. Future work could evaluate whether that additional effort yields sufficient gains in correctness, consistency, or downstream usability to justify the cost.

The results suggest that the LTL formulas themselves were often stronger than the wording of the propositions that supported them. In other words, even when the formula structure was largely correct, the proposition descriptions could still be poorly phrased, overly complex, or insufficiently reusable. For practical deployment, it would therefore be advisable either to define propositions ahead of time or to revise proposition wording after generation and before using the result in downstream formal verification tasks.

Finally, developers should pay particular attention to ambiguous requirements and to any generated formula that uses implication. Ambiguous requirements remain a source of interpretive variation even when overall pass@k performance is high. Similarly, implication should be checked carefully for complex requirements to ensure that the logical direction matches the intended outcome of the full requirement. Req 6 demonstrates that a formula can look reasonable while still being semantically inadequate. Consequently, implication-heavy translations should be manually inspected with special care. These recommendations are summarized by Table V.

### D. *Understanding Through Explanations and ltl2timeline*

Each LLM generation included one or two paragraphs briefly explaining why the generated LTL matches the NL requirement. These explanations were particularly helpful for difficult translations or hard-to-understand LTLs. However, LLM-generated descriptions may not be enough for providing understanding to a non-expert developer. Therefore, it is recommended that organizations leverage existing tools for LTL visualization, specifically the ltl2timeline tool [9]. Ltl2timeline provides an accessible timeline-based visualization of any input LTL in a novel visualization format and was specifically curated to allow a developer who is a non-expert in formal specifications to better understand a target specification. Utilizing such a tool would allow the developer to cross-check the LLM generation and determine if the LTL adequately captures the NL requirement. Fig. 2 displays an example ltl2timeline output for the LTL '(f | !a) -> Xe' generated by GPT 5.2 for Req 1.

### E. *Broader Implications Beyond Specific Models*

Although the study evaluates six named LLMs, the significance of the results is not limited to these exact models. The pace of LLM development is such that specific model versions may become outdated faster than the research cycle can fully accommodate. For that reason, the central value of the findings lies less in the ranking of individual models and more in the behavioral patterns revealed across a representative set of modern systems.

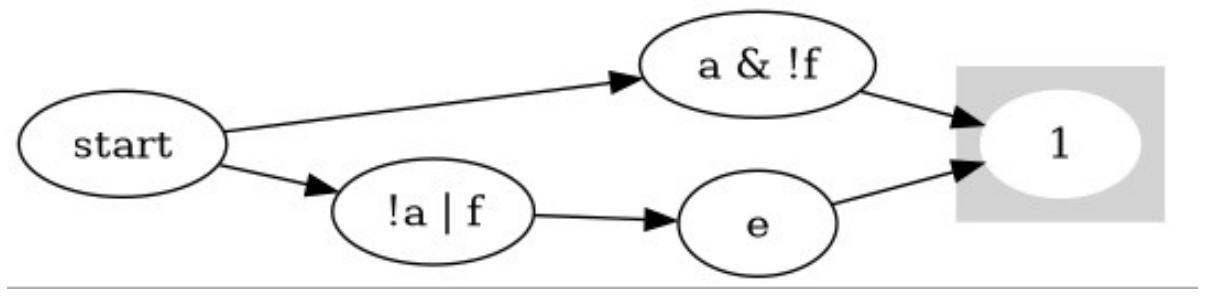

Fig. 2. Ltl2timeline output for the LTL '(f | !a) -> Xe'.

Several of those patterns are likely to remain informative even as model versions change. Structured requirements tend to be easier for LLMs to map into LTL than requirements that demand deeper reasoning about permissions, role asymmetries, or hidden logical constraints. Multiple samples can substantially improve the chance of obtaining a correct translation, but this benefit may be less valuable in settings where formal-methods expertise is scarce and pass@1 performance is therefore important. Proposition selection remains a persistent weakness, particularly with respect to atomicity and reuse. Finally, implication is sometimes a risky operator whose misuse can produce formulas that are syntactically well-formed yet semantically misleading.

These observations provide a snapshot of the current state of LLM capability on the unstructured NL-to-LTL task. Even if the exact models tested eventually become obsolete, the identified strengths and failure modes should still inform how newer models are evaluated and deployed. In that sense, the study serves not only as a model comparison, but also as an assessment of the maturity of contemporary general-purpose LLMs for this class of formalization task.

### *F. Extensions Beyond LTL*

The implications of this research also extend beyond LTL specifically. Although LTL is a natural target language because of its importance in reactive-system specification and model checking, the broader challenge addressed here is the automatic translation of unstructured requirements into formal artifacts. The same fundamental issues identified in this study—proposition selection, preservation of semantic scope, correct handling of implication and temporal structure, and the need for interpretable intermediate representations—are relevant to many other formal specification languages.

Accordingly, future work could investigate comparable workflows for other target formalisms, including branching-time logics, real-time temporal logics, contract-based specification languages, state-machine formalisms, and annotation languages used in software verification. LLMs may be similarly useful as front-end formalization assistants in these settings as well, provided that the prompting, validation, and review process are adapted to the semantics of the target language. In this broader context, this study suggests that modern LLMs are becoming viable components of semi-automated formalization pipelines more generally.

## VII. CONCLUSION

The results indicate that contemporary off-the-shelf LLMs can perform the translation of unstructured NL requirements into LTL with a level of accuracy that is already encouraging under few-shot prompting alone. In particular, the models achieved notable pass@1 performance and generally strong pass@3 and pass@5 results, suggesting that correct or near-correct formalizations can often be obtained within a small number of attempts even without task-specific fine-tuning. These findings are especially significant given the semantic variability and ambiguity inherent in unstructured requirements. In addition to translation accuracy, the generated NL explanations improved the understandability of the resulting formulas by making the relationships between requirement phrases and LTL structure more explicit. Such explanations can help clarify model assumptions and support manual review by developers with limited formal-methods expertise. To further strengthen this workflow, ltl2timeline should be leveraged alongside LLM-generated explanations, as timeline-based visualization can provide an additional and more intuitive representation of temporal behavior [9].

TABLE V. FINDINGS AND KEY TAKEAWAYS

| No. | Finding | Takeaway |
|---|---|---|
| 1 | 11/15 requirements approached or reached perfect performance. | Off-the-shelf LLMs are effective at many NL-to-LTL tasks. |
| 2 | LLMs performed well on requirements that clearly expressed condition-response and temporal scope. | The more structured the requirement is, the more likely the translation task is to succeed. |
| 3 | On Req 6, LLMs consistently misused implication to result in logic that contradicts the original NL. | Complex requirements utilizing implication should be human checked. |
| 4 | On Req 8, LLMs inadequately represented role-sensitive logic. | LTLs specifying role-specific logic should be human checked. |
| 5 | Gemini 3 Pro and Claude Opus 4.5 achieved the highest consistency score. | Consistency is not related to performance. |
| 6 | LLMs displayed difficulty correctly formulating propositions. | LLMs should be guided in formulating atomic and useful propositions. |

More broadly, continued investigation into NLP and LLM-based methods for automatically translating unstructured NL into formal specifications has the potential to increase industrial adoption of formal methods by reducing the time, specialized expertise, and therefore overall cost required to produce formal specifications. At the same time, the findings should not be interpreted as specific to any one model family or version; rather, they provide a snapshot of the current state of off-the-shelf LLM capability on the unstructured-requirement-to-LTL task and highlight patterns that are likely to remain relevant as models evolve. The implications of the study also extend beyond LTL, as many of the same challenges and opportunities apply to other formal specification languages. Future research will examine whether translation correctness improves when the correct atomic propositions are provided in advance, thereby isolating temporal and logical composition from proposition generation, and will extend evaluation beyond LTL to additional formal specification languages and representations. Such directions would help determine the generality of the observed results and further clarify the role of LLMs as practical front-end assistants in semi-automated formalization workflows.

## VIII. DATA AVAILABILITY STATEMENT

The supplemental tables detailing the requirements as well as the collected data, data analysis, and data collection prompt are provided for review at the following repository:
https://doi.org/10.6084/m9.figshare.31807135

## Acknowledgment

This work is supported by the National Science Foundation under Grant No. 2445056. Any opinions, findings, and conclusions or recommendations expressed in this material are those of the authors and do not necessarily reflect the views of the National Science Foundation.

## References

[1] R. Lorch et al., "Formal Methods in Requirements Engineering: Survey and Future Directions," in *International Conference on Formal Methods in Software Engineering (FormaliSE)*, Lisbon, Portugal, 2024.

[2] I. Buzhinsky, "Formalization of Natural Language Requirements into Temporal Logics: A Survey," in *International Conference on Industrial Informatics (INDIN)*, Helsinki, Finland, 2019.

[3] M. Zhao, R. Tao, Y. Huang, J. Shi, S. Qin and Y. Yang, "NL2CTL: Automatic Generation of Formal Requirements Specifications via Large Language Models," in *Formal Methods and Software Engineering (ICFEM 2024)*, Hiroshima, Japan, 2024.

[4] J. P. Gibson, "Formal Requirements Engineering: Learning from the Students," in *Australian Software Engineering Conference*, Canberra, ACT, Australia, 2000.

[5] F. Zahid, A. Tanveer, M. M. Y. Kuo and R. Sinha, "A Systematic Mapping of Semi-Formal and Formal Methods in Requirements Engineering of Industrial Cyber-Physical Systems," *Journal of Intelligent Manufacturing,* vol. 33, pp. 1603-1638, 2022.

[6] M. A. Rosado da Cruz and E. F. Cruz, "Machine Learning Techniques for Requirements Engineering: A Comprehensive Literature Review," *Software,* vol. 4, no. 14, 2025.

[7] J. Woodcock, P. G. Larsen, J. Bicarregui and J. Fitzgerald, "Formal Methods: Practice and Experience," *ACM Computing Surveys (CSUR),* vol. 41, no. 4, 2009.

[8] P. Spoletini and A. Ferrari , "The Return of Formal Requirements Engineering in the Era of Large Language Models," in *Requirements Engineering: Foundation for Software Quality (REFSQ 2024)*, Winterthur, Switzerland, 2024.

[9] R. Li, K. Gurushankar, M. J. Heule and K. Y. Rozier, "What's in a Name? Linear Temporal Logic Literally Represents Time Lines," in *IEEE Working Conference on Software Visualization (VISSOFT)*, Bogotá, Colombia, 2023.

[10] R. Jhala and R. Majumdar, "Software Model Checking," *ACM Computing Surveys,* vol. 41, no. 4, 2009.

[11] K. Y. Rozier, "Linear Temporal Logic Symbolic Model Checking," *Computer Science Review,* vol. 5, no. 2, pp. 163-203, 2011.

[12] M. Cosler, C. Hahn, D. Mendoza, F. Schmitt and C. Trippel, "nl2spec: Interactively Translating Unstructured Natural Language to Temporal Logics with Large Language Models," in *Computer Aided Verification*, Paris, France, 2023.

[13] S. Ghosh, D. Elenius, W. Li, P. Lincoln, N. Shankar and W. Steiner, "ARSENAL: Automatic Requirements Specification Extraction from Natural Language," in *NASA Formal Methods*, Minneapolis, MN, USA, 2016.

[14] D. Giannakopoulou, T. Pressburger, A. Mavridou and J. Schumann, "Automated Formalization of Structured Natural Language Requirements," *Information and Software Technology,* vol. 137, 2021.

[15] I. T. Leong and R. Barbosa, "Translating Natural Language Requirements to Formal Specifications: A Study on GPT and Symbolic NLP," in *IEEE/IFIP International Conference on Dependable Systems and Networks Workshops (DSN-W)*, Porto, Portugal, 2023.

[16] J. Wang, "Temporal Logic," in *Formal Methods in Computer Science*, Boca Raton, FL, USA, Taylor & Francis Group, LLC, 2020, pp. 155-156.

[17] M. Chen, J. Tworek, H. Jun, Q. Yuan, H. P. d. O. Pinto and J. Kaplan, "Evaluating Large Language Models Trained on Code," *arXiv preprint arXiv:2107.03374,* 2021.

[18] J. L. Fleiss, "Measuring Nominal Scale Agreement Among Many Raters," *Psychological Bulletin,* vol. 76, no. 5, 1971.

[19] A. Ferrari, G. O. Spagnolo and S. Gnesi, "PURE: a Dataset of Public Requirements Documents," in *IEEE International Requirements Engineering Conference (RE)*, Lisbon, Portugal, 2017.

[20] OpenAI, "Update to GPT-5 System Card: GPT-5.2," 2025.

[21] OpenAI, "GPT-5 System Card," arXiv preprint arXiv:2601.03267, 2025.

[22] Anthropic, "System Card: Claude Opus 4.5," 2025.

[23] Google Deepmind, "Gemini 2.5: Pushing the Frontier with Advanced Reasoning, Multimodality, Long Context, and Next Generation Agentic Capabilities," 2025.

[24] Google Deepmind, "Gemini 3 Pro - Model Card," 2025.

[25] NVIDIA, "Llama-Nemotron: Efficient Reasoning Models," arXiv preprint arXiv:2505.00949, 2025.

[26] P. Sahoo, A. K. Singh, S. Saha, V. Jain, S. Mondal and A. Chadha, "A Systematic Survey of Prompt Engineering in Large Language Models: Techniques and Applications," arXiv preprint arXiv:2402.07927, 2024.